\documentclass[12pt]{iopart}

\usepackage{iopams}  
\pdfoutput=1
\usepackage[colorlinks,linkcolor=blue,citecolor=blue,anchorcolor=blue]{hyperref}   
\usepackage[mathscr]{euscript}
\usepackage{graphicx,color}

\begin{document}

\title
{A generalization of the quantum Rabi model: 
exact solution
and spectral structure}

\author{Hans-Peter Eckle$^1$ and Henrik Johannesson$^{2,3}$}

\address{$^1$ Humboldt Study Centre,
Ulm University, D--89069 Ulm, Germany}

\address{$^2$ Department of Physics, University of Gothenburg, 
SE 412 96 Gothenburg, Sweden}

\address{$^3$ Beijing Computational Science Research Center, Beijing 100094, China}

\vspace{10pt}

\begin{abstract}
We consider a generalization of the quantum Rabi model where the two-level system and 
the single-mode cavity oscillator
are coupled by an additional Stark-like term. 
By adapting a method recently introduced by Braak 
[Phys. Rev. Lett. {\bf 107}, 100401 (2011)],
we solve the model exactly. 
The low-lying spectrum in the experimentally relevant ultrastrong and deep strong regimes 
of the Rabi coupling
is found to exhibit two striking features absent from the original quantum Rabi model: 
avoided level crossings for states of the same parity 
and an anomalously rapid onset of two-fold near-degenerate levels as the Rabi coupling 
increases. \\ \\
{\bf See pages 26-27 for an added corrigendum with corrected equations (39) and (40), and new figures replacing figures 4 and 6. 
The general conclusions drawn in the paper remain unchanged and valid.}  
\end{abstract}

%
\vspace{2pc}
\noindent{\it Keywords}: quantum optics, quantum Rabi-Stark model, 
                                       regular and exceptional spectrum
%
%
%
%


\section{Introduction}
\label{section:Introduction}
The integration of coherent nanoscale systems with quantum resonators is a focal point
of current quantum engineering of states and devices. 
Examples range from trapped ions interacting with a cavity field \cite{Leibfried2003} 
to superconducting 
charge qubits in circuit QED architectures \cite{Blais2007}.
The paradigmatic model for these systems is the Rabi model \cite{Rabi}
which was first introduced 80 years ago 
to discuss the phenomenon of nuclear magnetic resonance in a semi--classical way.
While Rabi treated 
the atom 
quantum mechanically, he still construed the 
rapidly varying weak magnetic field 
as a rotating classical field \cite{AllenEberly}.

In the course of investigating the relationship between the quantum theory of radiation
and the corresponding semi-classical theory,
Jaynes and Cummings \cite{JC} discussed a model similar to Rabi's. 
However, their model of an idealized atom consisting of only 
two levels coupled to a single quantised oscillator
mode in an optical cavity was now a fully quantum mechanical model,
the quantum Rabi model (sometimes also designated as the quantum
electrodynamic Rabi problem \cite{AllenEberly}).
Jaynes and Cummings \cite{JC} also introduced an important
approximation to the quantum
Rabi model, the so--called rotating wave approximation (RWA), leading to a model which can
be solved exactly by elementary means and which now bears the
name quantum Jaynes--Cummings model. 

The quantum Rabi model, on the other hand,
although still describing the interaction between 
matter and light in one of the simplest ways, 
only recently yielded to an exact and complete analytical solution \cite{Braak2011} 
when Braak found an ingenious way to exploit the underlying $\mathbb{Z}_2$ parity
symmetry of the model to derive its energy spectrum.
While the quantum Jaynes--Cummings model has sufficed for a long time to describe
experiments in quantum optics, recently it has become more and more necessary to
go beyond the RWA as the larger Rabi coupling strengths 
of the ultrastrong and deep strong regimes
come within experimental reach \cite{NiemczykEtAl2010,CasanovaEtAl2010}. 

In connection with his investigation of the exact solvability of the 
quantum Rabi model, Braak also developed a
new proposal for quantum integrability \cite{Braak2011}.
This proposal is of considerable importance in view of the ongoing quest
for a consistent notion of quantum integrability
\cite{Weigert,CauxMossel,Larson,BatchelorZhou}.

Concurrent with this theoretical
breakthrough, and motivated mostly by novel experimental setups,
there has been an avalanche of studies of the quantum Rabi model 
and its many generalizations, revealing a plethora of intriguing and
intrinsically nonclassical effects (for a recent review, see \cite{BCReview}).

A particularly interesting generalization of the model was proposed by Grimsmo and 
Parkins in 2013 \cite{GP2013}. 
These authors inquired about the possibility to realize the quantum Rabi model with a
single atom coupled to a high--finesse optical cavity mode. 
They arrived at a scheme where two hyperfine ground states of
a multilevel atom emulate an effective two--level system, with resonant Raman transitions 
between the two states induced by the cavity field and two auxiliary laser fields. 
Importantly, this scheme allows for a realization of the quantum Rabi model where 
coupling constants and effective frequencies can be freely and independently tuned, 
opening an experimental inroad to systematically probe also the ultrastrong and 
deep--strong coupling regimes. These are the regimes where the Rabi model comes into its own, 
while the time--honoured RWA $-$ which 
allowed the Rabi model to be replaced by the much simpler Jaynes--Cummings model \cite{JC} $-$
breaks down. 

For generic values of the parameters of the model, 
however, the Grimsmo--Parkins
scheme requires the addition of a new term to  the quantum Rabi Hamiltonian, 
a nonlinear coupling term between the two--level system and the quantum oscillator.
Such a coupling term has been discussed in the quantum optics 
literature under the name of 
dynamical Stark shift, a quantum version of the Bloch--Siegert
shift \cite{Klimov2009}.
Accordingly, we shall call the quantum Rabi model augmented by a nonlinear term
of the kind discussed by Grimsmo and Parkins the quantum Rabi--Stark model.

Note, however, that in the usual dynamical Stark shift the corresponding nonlinear
coupling strength is determined by the parameters of the underlying quantum Rabi
model.
In the scheme proposed by Grimsmo and Parkins \cite{GP2013} also the Stark coupling
can be adjusted freely and independently.

Grimsmo and Parkins conjecture \cite{GP2014} that the Rabi--Stark model
may undergo a superradiant transition in the deep strong coupling regime
of the Rabi coupling
when the Stark coupling strength becomes equal to the frequency of the cavity 
mode.
The additional nonlinear term in the Hamiltonian, the Stark term, may therefore
give rise to new physics.
It will therefore be of considerable importance to thoroughly investigate the
spectral properties of the quantum Rabi--Stark model.

The exact solvability of the model has been elegantly demonstrated in recent work 
by Maciejewski {\it et al.} \cite{StarkPolish,StarkPolish2}, using a Bargmann
representation.
The ensuing coupled set of differential equations were then solved by a
technique involving Wronskian determinants in the general case
and an analysis based on the  
Stokes phenomenon \cite{Balser} for the special case when the Stark coupling
becomes equal to the quantum oscillator frequency.

In this paper we take a different route to obtain the exact solution of the quantum
Rabi--Stark model,  adapting Braak's method from 2011 \cite{Braak2011} developed
for the original quantum Rabi model. 
This alternative approach has the virtue of laying bare certain 
structural similarities between the two models, and highlights the 
importance of the underlying $\mathbb{Z}_2$ parity symmetry
which is present also in the quantum Rabi--Stark model. 
In particular $-$
according to Braak's criterion for quantum integrability \cite{Braak2011} $-$ the retaining
of the $\mathbb{Z}_2$ symmetry implies that also the Rabi-Stark model is integrable.

Almost all energy eigenvalues are determined by the zeros of two transcendental 
functions, obtained from a Frobenius analysis of the coupled singular differential 
equations which define the eigenvalue problem in the Bargmann
representation. 
Provided that the model parameters are chosen so that these transcendental 
functions are reasonably well--behaved, this allows for 
numerical access to large portions of the spectrum. 
Fortunately, the parameter regimes where this property holds cover the most 
interesting cases for current experiments: the ultrastrong and opening deep strong 
regimes of the Rabi coupling. 

There also exist, again like in the original quantum Rabi model, 
exceptional 
spectral points which do not correspond to zeros 
of these transcendental functions, 
but to points in parameter space where the 
singularities of the transcendental functions are lifted. 
As for the original quantum Rabi model, the exceptional solutions 
may define level crossings in the spectrum between energy levels of different parity. 
By increasing the magnitude of the 
Stark coupling
we find 
that these level crossing points 
become 
less and less frequent.
Instead there is a stronger tendency $-$ as compared to the original quantum Rabi model 
$-$  for neighboring levels to coalesce and eventually become
two-fold degenerate. 
This surprising effect comes about from a ``reshuffling" of energy levels caused by 
the added nonlinear Stark coupling, yielding a compressed spectrum which 
favors pairwise degenerate levels as the two--level system gets coupled to the 
quantum oscillator more strongly.   

The layout of the paper is as follows: 
In the next section, section \ref{section:model}, 
we introduce the model, with reference to \cite{GP2013}, 
and discuss some of its key properties. 
Section \ref{section:solution}
contains the analytical solution of the model, leading up to the construction 
of the transcendental functions, the zeros (lifted singularities) of which determine
the regular (two-fold degenerate exceptional) part of the exact spectrum
(which becomes complete when adding also 
the non-degenerate exceptional part of the spectrum, 
as discussed in section \ref{section:solution}).
In section \ref{section:spectrum}, the spectral structure
in the ultrastrong and opening deep strong coupling 
regimes is extracted numerically from the exact solution, and the novel 
features $-$ as compared to that of the original quantum Rabi model $-$ are highlighted 
and discussed. 
Section \ref{section:DS}, finally, contains a summary and outlook.
 
\section{The quantum Rabi--Stark model}
\label{section:model}

As we have expounded in the introduction,
the quantum Rabi model describes the interaction between light and matter,
next to the Jaynes--Cummings model, in
the simplest possible way and is used as a basic model in many 
fields of physics
\cite{HarocheRaimond2006}. 

The simplest generic experimental set--up to realize the quantum Rabi model,
a cavity quantum electrodynamics (cavity QED) system, 
consists of a single atom 
put into a single--mode photon field 
which is enclosed by mirrors in a cavity. 
The frequency of the single--mode photon is chosen in such a way as to 
interact predominantly only with two levels of the atom \cite{Walther2006}

In an experiment, there will inevitably be processes which lead to dissipative losses.
In a cavity QED experiment, such processes include the 
dissipative loss of photons from the cavity (at rate $\kappa$) and the
emission of the atom into other modes than the single cavity mode (at rate $\tau$).
If such losses can be made small compared to the interaction strength 
between
the single photon mode and the atom, described now as a two--level system,
the experimental situation can be described by the quantum Rabi model Hamiltonian
\begin{eqnarray}
\label{eq:RH1}
{\mathcal H}_\mathrm{Rabi}&=&\omega a^\dagger a+\Delta\sigma^z+
g\sigma^x(a+a^\dagger)\\
\label{eq:RH2}
&=&\omega a^\dagger a+\Delta\sigma^z+
g\left(\sigma^++\sigma^-\right)(a+a^\dagger),
\end{eqnarray}
where $a^\dagger$ and $a$ are the creation and annihilation operators of the
quantum oscillator mode with frequency $\omega$.
The two--level atom is described by the Pauli matrices $\sigma^x$ and $\sigma^z$
with the splitting between the two levels given by $\Delta$.
The interaction strength between
the single photon mode and the two--level system is $g$ which we call the Rabi coupling
to distinguish it from the Stark coupling which will be introduced below.

As already mentioned in the introduction,
the Rabi model was originally introduced as the basis to understand 
nuclear magnetic resonance \cite{Rabi} 
and has since been applied to physical systems ranging from quantum
optics to condensed matter physics, e.g.\ cavity and circuit 
quantum electrodynamics, quantum dots, trapped ions, and superconducting 
qubits. Moreover, it is used to 
describe nanoelectromechanical
devices where the role of the photons is taken by phonons (see, for instance,
\cite{GellerCleland2005} and \cite{Tian2011}). 
These physical systems are also under investigation
as candidates for the physical realization of quantum information processing.

Grimsmo and Parkins \cite{GP2013}
propose an experimental arrangement where the two
relevant levels of a $^{87}$Rb atom in the single--mode cavity is subjected 
to two auxiliary laser beams.
Under conditions equivalent to the ones described above where losses can be
neglected, Grimsmo and Parkins can describe their proposed experimental
arrangement by an effective Hamiltonian
\begin{eqnarray}
\label{eq:RSH1}
{\mathcal H}&=&{\mathcal H}_\mathrm{Rabi}+\gamma\sigma^za^\dagger a\\
\label{eq:RSH2}
&=&\omega a^\dagger a+\Delta\sigma^z+
g\sigma^x(a+a^\dagger)+\gamma\sigma^za^\dagger a,
\end{eqnarray}
where an additional term, $\gamma\sigma^za^\dagger a$, appears compared to the 
original quantum Rabi Hamiltonian ${\mathcal H}_\mathrm{Rabi}$.
This  additional term models a nonlinear coupling 
between the two--level atom and the single--mode cavity oscillator.
In the introduction, we gave an argument for naming this Hamilton and the corresponding
model the quantum Rabi--Stark Hamiltonian and model, respectively, with the
coupling constant $\gamma$, the Stark coupling.

The Hamiltonian (\ref{eq:RH2}) of the original quantum Rabi model is solvable by  
elementary means, employing the RWA
(see for example \cite{AllenEberly} where also the classical and
semi--classical versions of the Rabi model are discussed).
The resulting model, the quantum Jaynes--Cummings model, emerges through 
the RWA 
by neglecting the terms $a^\dagger\sigma^+$ and $a\sigma^-$ in
the Hamiltonian (\ref{eq:RH2}).

The Jaynes--Cummings model can also be investigated
with an analogous nonlinear Stark term added.
Interestingly, this variant of the Jaynes--Cummings model sheds light on the
Bethe ansatz solution of the original Jaynes--Cummings model.
The former can be solved by a standard algebraic Bethe ansatz procedure which allows
to extract the algebraic solution of the latter in the limit 
when the Stark term vanishes
\cite{BBT,Book}.

\section{Exact solution of the Rabi-Stark model}
\label{section:solution}
In this section, we shall outline the exact solution of the 
quantum Rabi--Stark model represented 
by the Hamiltonian (\ref{eq:RSH2}).
In doing so, we shall generalize the method introduced by Braak in \cite{Braak2011}
for the solution of the original quantum Rabi model, described by the Hamiltonian
(\ref{eq:RH1}), and especially highlight those aspects where the two models differ.

\subsection{Bargmann space representation of the eigenvalue problem}
\label{section:Bargmann}
It will prove advantageous to rewrite the quantum Rabi--Stark
Hamiltonian (\ref{eq:RSH2})
in the spin--Boson representation, achieved through a unitary rotation
of the Hamiltonian by the operator
$e^{i\pi\sigma^y/4}$.
The Hamiltonian (\ref{eq:RSH2}) then becomes
\begin{equation}
\label{eq:RSHBoson}
{\mathcal H}=\omega a^\dagger a+\Delta\sigma^x+g\sigma^z(a+a^\dagger)
+\gamma\sigma^x a^\dagger a.
\end{equation}
In order to calculate the eigenvalues of this Hamiltonian exactly,
we employ the Bargmann space representation \cite{Bargmann} 
(for a recent summary,
with a view on its application to the quantum Rabi model,
of the properties of the Bargmann space representation, which is isomorphic to the
space of square integrable functions ${\mathscr L}^2(\mathbb{R})$,
see \cite{BraakReview}).
In the Bargmann space representation, the quantum oscillator 
creation operator is replaced by a 
complex variable $z$, i.e.\ $a^\dagger\rightarrow z$, and the quantum oscillator
annihilation operator by the derivative with respect to the complex variable, i.e.\
$a\rightarrow d/dz$.
The
state vector $|\psi\rangle$ is represented in the Bargmann space
representation by a wave function 
$u(z)$
depending on the complex variable $z$.

We briefly state the two requirements a function 
$u(z)$ needs to satisfy
in order 
to be an admissible function of the Bargmann space $\mathscr{B}$, i.e.\
to be a physically allowed wave function. 
These requirements were carried over by Bargmann
from the corresponding requirements which wave
functions have to satisfy in the space of square integrable functions
${\mathscr L}^2(\mathbb{R})$.
The first requirement is that the function must have a finite norm 
$\langle u|u\rangle<\infty$, where the scalar product is defined by
\begin{equation}
\langle u|v\rangle=\frac{1}{\pi}\int_\mathbb{C}\,d\Re(z)\,d\Im(z)\,
\overline{u(z)}v(z)e^{-z\bar{z}},
\end{equation}
and the second requirement that it 
be holomorphic everywhere in $\mathbb{C}$, i.e.\ be an entire function
\cite{Bargmann}.

Measuring energy in units of the quantum oscillator 
frequency, i.e.\ formally putting $\omega=1$,
the Rabi--Stark Hamiltonian 
(\ref{eq:RSHBoson}) becomes in the Bargmann representation
\begin{equation}
\label{eq:HNonDiag}
{\mathcal H}=
\left(
\begin{array}{cc}
z\frac{d}{dz}+g\left(z+\frac{d}{dz}\right)  & \gamma z\frac{d}{dz}+\Delta   \\
\gamma z\frac{d}{dz}+\Delta  & z\frac{d}{dz}-g\left(z+\frac{d}{dz}\right)
\end{array}
\right).
\end{equation}
The canonical Fulton--Gouterman transformation \cite{FG}
\begin{equation}
{\mathcal U}=\frac{1}{\sqrt{2}}
\left(
\begin{array}{cc}
1  & 1  \\
{\mathcal T} &   -{\mathcal T}
\end{array}
\right),
\end{equation}
employing the parity operator ${\mathcal T}[u(z)]=u(-z)$,
transforms the Hamiltonian (\ref{eq:HNonDiag}) onto diagonal form
\begin{equation}
{\mathcal U}^{-1}{\mathcal H}\,{\mathcal U}=
\left(
\begin{array}{cc}
{\mathcal H}_+   & 0  \\
 0  &  {\mathcal H}_- 
\end{array}
\right)
\end{equation}
with the Hamiltonians
\begin{equation}
{\mathcal H}_\pm=z\frac{d}{dz}+g\left(z+\frac{d}{dz}\right)
\pm\left(\gamma z\frac{d}{dz}+\Delta\right){\mathcal T}
\end{equation}
in the parity Hilbert spaces $\mathscr{H}_\pm$.
The corresponding Schr\"{o}dinger equations in the 
positive and
negative parity sectors, respectively,
\begin{equation}
\label{eq:SEq}
{\mathcal H}_\pm\psi^{(\pm)}(z)=E_\pm\psi^{(\pm)}(z)
\end{equation}
become, written explicitly, non--local functional differential equations
\begin{equation}
\label{eq:FDE}
\!\!\!\!\!\!\!\!\!\!\!\!
z\frac{d}{dz}\psi^{(\pm)}(z)+g\left(\!z+\frac{d}{dz}\!\right)\psi^{(\pm)}(z)
\pm\left(\!\gamma z\frac{d}{dz}+\Delta\!\right)\psi^{(\pm)}(-z)=E_\pm\psi^{(\pm)}(z).
\end{equation}
These two differential equations are converted into each other by the simultaneous
replacements  $\gamma\rightarrow-\gamma$ and $\Delta\rightarrow-\Delta$.
It is therefore sufficient, and we shall do this in the following, to concentrate on
one differential equation, here chosen as the one in the positive parity sector.  

The reducibility of the Bargmann representation (\ref{eq:HNonDiag}),
into two blocks ${\mathcal H}_{\pm}$ with definite parities $\pm 1$, 
reflects that the Rabi-Stark
Hamiltonian (\ref{eq:RSH1}) is invariant under the $\mathbb{Z}_2$ parity transformation
\begin{equation}
\label{eq:ParityOp}
{\mathcal P}=(-1)^{a^\dagger a}\sigma^z.
\end{equation}
Hence, the eigenstates $|\psi\rangle$ can be labeled by the energy eigenvalue
$E$ and the parity eigenvalue $p=\pm1$,
\begin{equation}
|\psi\rangle=|E,p\rangle.
\end{equation}
The $\mathbb{Z}_2$ parity symmetry is crucial for both, 
the exact solution of the model, and
also its quantum integrability according to the quantum integrability
criterion proposed by Braak
\cite{Braak2011}.

Returning to (\ref{eq:FDE}), in order to deal with the non--locality of the differential equation for $\psi^{(+)}$,
we define the two new functions (dropping the upper index $(+)$ 
for the time being)
\begin{equation}
\label{eq:SF}
\phi(z)\equiv\psi(z)\qquad\mbox{and}\qquad\bar{\phi}(z)\equiv\psi(-z),
\end{equation}
thus obtaining a set of two local differential equations. 
Note that this definition means that we now have two representations of the same
function $\psi(z)$ which are to be determined from the two coupled local differential
equations.
With these definitions and rearranging terms, this set of two coupled
local differential equations becomes explicitly
\begin{eqnarray}
\label{eq:ode1}
(z+g)\frac{d}{dz}\phi(z)+\left(gz-E\right)\phi(z)
+\gamma z\frac{d}{dz}\bar{\phi}(z)+\Delta\bar{\phi}(z)&=&0,\\
\label{eq:ode2}
(z-g)\frac{d}{dz}\bar{\phi}(z)-\left(gz+E\right)\bar{\phi}(z)
+\gamma z\frac{d}{dz}\phi(z)+\Delta\phi(z)&=&0.
\end{eqnarray}
Note that these two first--order complex differential equations are coupled in both,
the unknown functions $\phi(z)$ and $\bar{\phi}(z)$ and their derivatives
$d\phi(z)/dz$ and $d\bar{\phi}(z)/dz$.
This is an important difference and complication compared to the original quantum Rabi model
and is due to the nonlinear term proportional to the 
Stark coupling strength $\gamma$.

\begin{figure}
\begin{center}
\includegraphics[width=.65\textwidth]{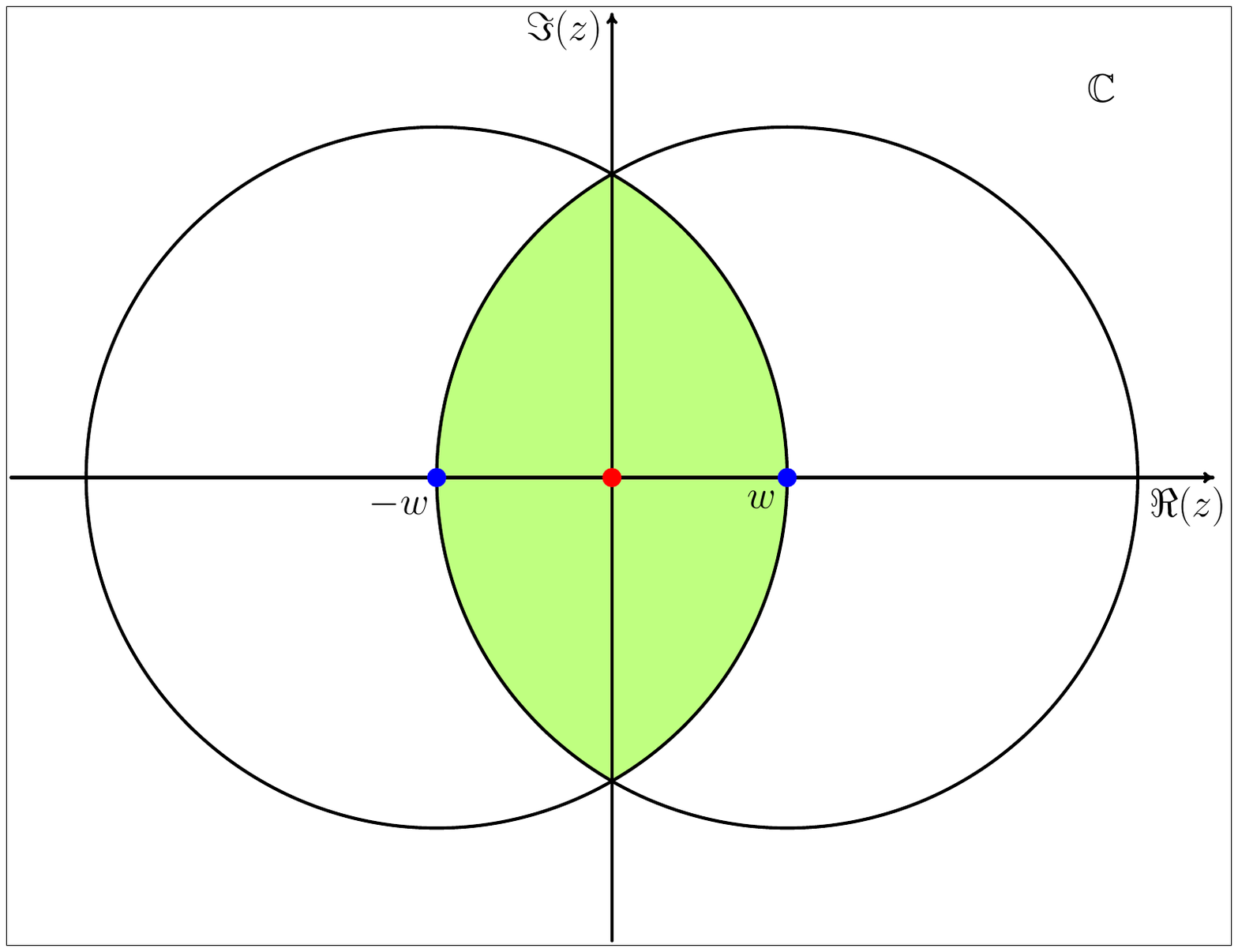}
\caption{Singularity structure of the differential equations 
              (\ref{eq:ODEa}) and (\ref{eq:ODEb}).
                  The regular singular points are at $\Re{z}=\pm w, \Im{z}=0$ 
                  (blue dots) with 
                  $w=g/\sqrt{1-\gamma^2}$,
                  the irregular singular point is at $z=\infty$.
                  While all other points of the complex plane $\mathbb{C}$ are ordinary
                  points, the ordinary point at $\Re{z}=\Im{z}=0$ 
                  (red dot) will play a
                  particularly prominent role in obtaining the spectrum, cf.\ sections 
                  \ref{section:solution} and \ref{section:spectrum}.
                 } 
\label{fig:GammaDisksV3}
\end{center}
\end{figure}

The two coupled first--order differential equations can be 
partially decoupled with respect to the
coupling of the derivatives.
In compact notation, we obtain the set of two first--order ordinary differential equations
\begin{eqnarray}
\label{eq:ODEa}
\Gamma(z)\phi^\prime&=&\Lambda(z)\phi-\bar{{\mathcal E}}(z)\bar{\phi},
\\
\label{eq:ODEb}
\Gamma(z)\bar{\phi}^\prime&=&\bar{\Lambda}(z)\bar{\phi}-{\mathcal E}(z)\phi,
\end{eqnarray}
where we introduced the functions
\begin{equation}
\label{eq:Gamma}
\Gamma(z)=(1-\gamma^2)(z-w)(z+w)
\end{equation}
with
$w=g/\sqrt{1-\gamma^2}$,
and
\begin{eqnarray}
\Lambda(z)&=&(E-gz)(z-g)+\gamma\Delta z, \ \ \bar{\Lambda}(z)=(E+gz)(z+g)+\gamma\Delta z, 
\\
{\mathcal E}(z)&=&\Delta(z+g)+\gamma z(E-gz), \ \ \bar{{\mathcal E}}(z)=\Delta(z-g)+\gamma z(E+gz).
\end{eqnarray}
From these functions, especially (\ref{eq:Gamma}), we observe
that the differential equations 
are singular with regular singularities at $z=\pm w$ (see figure \ref{fig:GammaDisksV3}).
Note that the regular singularities in the Rabi--Stark model depend on both,
the Rabi coupling $g$ and the Stark coupling $\gamma$.

Furthermore, the equations have an irregular singularity at $z=\infty$ of 
s--rank $R(\infty)=2$ \cite{Slavyanov} which
can be demonstrated by transforming the equations into second--order equations outside
of a sufficiently large disk of radius
$|z|=R$ which includes all singularities lying in a finite region of 
the complex  plane.
The s--rank $R(\infty)=2$ of the differential equations 
guarantees that the solutions have a finite norm asymptotically for $z\rightarrow\infty$  
and are thus members of the Bargmann space \cite{Slavyanov,Ince,BraakReview}.

\subsection{Frobenius analysis of the singular differential equations}
\label{section:Frobenius}
An indicial analysis \cite{Ince} of the Frobenius ansatz around the regular singular points
$z_0=\pm w$ 
\begin{equation} 
\phi
(z)
=\sum_{n=0}^\infty 
A_n
(z-z_0)^{n+r} 
\label{eq:SE1}
\end{equation}
of the \emph{decoupled} 
second--order differential equation for $\phi$ $-$ obtained from the coupled first-order differential equations in (\ref{eq:ODEa}) and (\ref{eq:ODEb}) $-$
reveals that there is
one indicial exponent 
\begin{equation}
\label{eq:xgamma}
r=r_1=\frac{E+g^2+\Delta\gamma}{1-\gamma^2}\equiv x_\gamma\geq0
\end{equation}
at each of the regular singular points $z_0=\pm w$.

The other indicial exponent is given by
\begin{equation}
r=r_2=0,
\end{equation}
again at both regular singular points $z_0=\pm w$.
The same indicial exponents are also obtained for the Frobenius ansatz
\begin{equation}
\bar{\phi}
(z)
=\sum_{n=0}^\infty 
\bar{A}_n
(z-z_0)^{n+r} 
\label{eq:SE2} 
\end{equation} 
from an indicial analysis of the second--order differential equation for
$\bar{\phi}$, again at both regular singular points $z_0=\pm w$.

There is a subtle point to note about the indicial analysis.
The limit $\gamma\rightarrow0$ does not in general reproduce the indicial exponents of 
the differential equations for the original quantum Rabi model \cite{BraakReview}.
The reason for this is that the indical analysis requires a limit $z\rightarrow\pm w$
which cannot be interchanged with the limit $\gamma\rightarrow0$.

It is important to stress that the indicial exponents determine whether the series 
solutions of the differential equations
are also physically acceptable solutions, i.e.\ wave functions, belonging to
the Bargmann space $\mathscr{B}$.
If $r\in\mathbb{N}_0$, 
this is the case. 
However, solutions for generic $r$, i.e.\ for values of $r\notin\mathbb{N}_0$,
although mathematically valid, are not members of the Bargmann space of physical wave
functions.

For the quantum Rabi model where $\gamma=0$, the further analysis of our
differential equations 
(\ref{eq:ODEa}) and (\ref{eq:ODEb}) can proceed directly \cite{BraakReview} 
or after transforming them into second--order equations \cite{BC1}.
In the present case, the transformation to second--order differential equations
generates further singularities not present
in the first--order equations which make the analysis difficult.
It is therefore preferable to directly solve the first--order equations as we shall do in
the following.
For generic values of the 
parameters $\{\Delta,\gamma,g\}$
and the energy eigenvalue $E$, the indicial
exponent $r_1$ will be a positive non--integer real number and, hence, 
the corresponding Frobenius
solution, exhibiting a branch cut,
will not be a member of the Bargmann space $\mathscr{B}$, i.e.\ will not be
a physical solution.
In these cases only the indicial exponents $r_2=0$ correspond 
to physical solutions 
$\phi(z)$ and $\bar{\phi}(z)$ belonging
to the Bargmann space.
The corresponding energy eigenvalues 
constitute the regular spectrum 
\cite{Judd,Kus,Tomka} of the Rabi--Stark model Hamiltonian.  

However, for special combinations of the parameters $\{\Delta,\gamma,g\}$ and the
energy eigenvalue $E$, the indicial exponent $r_1$ may become a non--negative
integer.
Such combinations give rise to the exceptional spectrum 
of the model, in close analogy with how exceptional spectra emerge
in Jahn-Teller-like systems, first discussed by Judd \cite{Judd}.

In the following section \ref{section:RegularSpectrum}, 
we concentrate our attention on the regular spectrum, while we shall discuss
the exceptional spectrum in section \ref{section:Espectrum}.

\subsection{Regular spectrum}
\label{section:RegularSpectrum}
Through the solution of the set of coupled
differential equations (\ref{eq:ODEa}) and (\ref{eq:ODEb}) for the case $r_2=0$, we obtain the
regular part of the spectrum.
We focus on the singularity at $z_0=-w$ and introduce the 
new complex variable $y=z-z_0=z+w$ to
perform a 
transformation of the functions $\phi(z)$ and $\bar{\phi}(z)$ according to
\begin{eqnarray}
\label{eq:T}
\phi(z)&=&e^{-wz}\rho(z)=e^{-wy+w^2}\rho(y),
\\
\label{eq:Tbar}
\bar{\phi}(z)&=&e^{-w z}\bar{\rho}(z)=e^{-w y+w^2}\bar{\rho}(y),
\end{eqnarray}
which implies for the first derivatives
\begin{eqnarray}
\frac{d\phi(z)}{dz}&=&e^{-w y+w^2}
\left(\frac{d}{dy}-w\right)\rho(y),
\\
\frac{d\bar{\phi}(z)}{dz}&=&e^{-w y+w^2}
\left(\frac{d}{dy}-w\right)\bar{\rho}(y),
\end{eqnarray}
such that 
the two first--order differential equations become
\begin{eqnarray}
(1-\gamma^2)(y-2w)y\rho^\prime&=&(K_2y^2+K_1y+K_0)\rho+
(\bar{K}_2y^2+\bar{K}_1y+\bar{K}_0)\bar{\rho},
\\
(1-\gamma^2)(y-2w)y\bar{\rho}^\prime&=&
(\bar{C}_2y^2+\bar{C}_1y+\bar{C}_0)\bar{\rho}+
(C_2y^2+C_1y+C_0)\rho
\end{eqnarray}
with the constants $K_2,\ldots, C_0$ depending on the parameters
$\{\Delta,\gamma,g\}$ and the energy eigenvalue $E$:
\begin{eqnarray}
K_2&=&(1-\gamma^2)w-g, 
\\
K_1&=&E-g^2+2gw+\gamma\Delta,
\\
K_0&=&-\left[(E+gw)(w+g)+\gamma\Delta w\right],
\\
\bar{K}_2&=&-\gamma g,
\\
\bar{K}_1&=&-\left[\Delta+\gamma(E-2gw)\right],
\\
\bar{K}_0&=&\Delta(w+g)+\gamma w(E-gw),
\end{eqnarray}
and \cite{Corrigendum},
\begin{eqnarray}
\bar{C}_2&=&2g,
\\
\bar{C}_1&=&E+g^2-4gw+\gamma\Delta,
\\
\bar{C}_0&=&-\left[(E-gw)(w-g)+\gamma\Delta w\right],
\\
C_2&=&\gamma g,
\\
C_1&=&-\left[\gamma(E+2gw)+\Delta\right],
\\
C_0&=&\gamma w(E+gw)+\Delta(w-g).
\end{eqnarray}
Writing $\rho(y)$ and $\bar{\rho}(y)$ as a power series
\begin{eqnarray}
\label{eq:L}
\rho(y)&=&\sum_{n=0}^\infty\alpha_n y^n,
\\
\label{eq:Lbar}
\bar{\rho}(y)&=&\sum_{n=0}^\infty\bar{\alpha}_n y^n,
\end{eqnarray}
where the expansion coefficients $\alpha_n$ and $\bar{\alpha}_n$ depend on the
parameters $\{\Delta,\gamma,g\}$ and the energy eigenvalue $E$,
we obtain a set of two coupled recursion relations for $n\geq2$,
\begin{eqnarray}
\nonumber
\!\!\!\!\!\!\!\!\!\!-K_2\alpha_{n-2}+\left((1-\gamma^2)(n-1)-K_1\right)\alpha_{n-1}
-\left(2w(1-\gamma^2)n+K_0\right)\alpha_n
&=&
\\
\label{eq:rra}
\bar{K}_2\bar{\alpha}_{n-2}+\bar{K}_1\bar{\alpha}_{n-1}+
\bar{K}_0\bar{\alpha}_{n},
\\
\nonumber
\!\!\!\!\!\!\!\!\!\!-\bar{C}_2\bar{\alpha}_{n-2}+
\left((1-\gamma^2)(n-1)-\bar{C}_1\right)\bar{\alpha}_{n-1}
-\left(2w(1-\gamma^2\right)n+\bar{C}_0)\bar{\alpha}_n
&=&
\\
\label{eq:rrb}
C_2\alpha_{n-2}+C_1\alpha_{n-1}+
C_0\alpha_{n}.
\end{eqnarray}
The recursion relations for $n=0$ and $n=1$ can be obtained directly
but also by the formal requirement that the 
expansion coefficients $\alpha_n$ and $\bar{\alpha}_n$
with index $n=-2$ and $n=-1$ vanish in the recursion relations
(\ref{eq:rra}) and (\ref{eq:rrb}).

For $n=0$, we obtain
\begin{eqnarray}
\label{eq:Ini1}
K_0\alpha_0+\bar{K}_0\bar{\alpha}_0&=&0,
\\
\label{eq:Ini2}
C_0\alpha_0+\bar{C}_0\bar{\alpha}_0&=&0,
\end{eqnarray}
i.e.\ a set of two homogeneous algebraic equations for $\alpha_0$ and $\bar{\alpha}_0$.
These algebraic equations have a non--trivial solution only if the coefficient determinant
vanishes, 
\begin{equation}
K_0\bar{C}_0-\bar{K}_0C_0=0.
\end{equation}
This determinant indeed vanishes identically for all values of the 
parameters $\{\Delta,\gamma,g\}$ and all values of the energy eigenvalue $E$.
The solutions of (\ref{eq:Ini1}) and (\ref{eq:Ini2}), 
\begin{eqnarray}
\alpha_0&=&-\frac{\bar{K}_0}{K_0}=-\frac{\bar{C}_0}{C_0},\\
\bar{\alpha}_0&=&1,
\end{eqnarray}
can therefore be used as initial
values for the coupled recursion relations (\ref{eq:rra}) and (\ref{eq:rrb}).

With the procedure described above,
we have now obtained the holomorphic solutions 
$\phi(z)$ and $\bar{\phi}(z)$ at the regular singular point $z_0=-w$
of the coupled set of the two first--order ordinary differential equations
(\ref{eq:ODEa}) and (\ref{eq:ODEb}).
These solutions are valid
in a disk of convergence of radius $2w$
around the regular singular point $z_0=-w$ (see figure \ref{fig:GammaDisksV3}).
They will, however, in general, i.e.\ for arbitrary values of the energy
eigenvalue $E$ not be holomorphic at the other regular singular point, $z_0=w$,
but will develop branch cuts at this singular point. 

On the other hand, by a corresponding analysis we can find the holomorphic 
solutions 
$\phi(z)$ and $\bar{\phi}(z)$
to (\ref{eq:ODEa}) and (\ref{eq:ODEb}) which are valid
in a disk of convergence of radius $2w$ around the regular singular point $z_0=w$. 
Again, these expansions, holomorphic at the regular singular point $z_0=w$,
will in general not be holomorphic at the other regular singular point $z_0=-w$.

The symmetry of the differential equations (\ref{eq:ODEa}) and (\ref{eq:ODEb})
under reflection $z\rightarrow-z$ reveals that the two combinations, 
written in vector notation as
$\left(\phi(z),\bar{\phi}(z)\right)^T$
and $\left(\bar{\phi}(-z),\phi(-z)\right)^T$,
satisfy the set of differential equations (\ref{eq:ODEa}) and (\ref{eq:ODEb}).
This property implies that, having obtained a holomorphic solution at one regular
singularity through the procedure outlined above, say at $z_0=-w$, 
we also have one at the other regular singularity, i.e.\ at $z_0=w$.
However, they represent one and the same function, as required in (\ref{eq:SF}),
only if the corresponding
energy eigenvalue $E$ belongs to the discrete spectrum of the Hamiltonian
(\ref{eq:RSH2}).
Then these solutions can serve as analytic continuations of each other.
Together with the s--rank $R(\infty)=2$ 
for the irregular singularity at $z\rightarrow\infty$, this
guarantees that we can find solutions of (\ref{eq:ODEa}) and (\ref{eq:ODEb}) which
satisfy the requirements for physical solutions of the Bargmann space $\mathscr{B}$.

In practice, the coupled recursion relations can only be solved numerically.
Assuming that we have obtained the expansion coefficients, at least to a sufficient degree
of numerical accuracy, we can extract the energy eigenvalue $E$ 
from the solutions of the first--order differential equations, i.e.\ the wave functions in the
Bargmann space representation.
This is done by adapting the $G_\pm$ function formalism 
developed by Braak \cite{Braak2011} for the quantum Rabi model
to our purposes of the generalization of the Rabi model, the quantum Rabi--Stark
model.
Reintroducing the parity label $(\pm)$ for the wave functions $\phi$ and $\bar{\phi}$, 
we accordingly
introduce the $G_\pm$ functions which are 
functions of the energy eigenvalue $E$, the parameters of the Hamiltonian
$\{\Delta,\gamma,g\}$, measured in units of the quantum oscillator frequency $\omega$, 
and the complex variable $z$
\begin{equation}
\label{eq:G}
G_\pm(
\pm\Delta,\pm\gamma,g|E;z)=\bar{\phi}^{(\pm)}
(-z)-\phi^{(\pm)}
(z).
\end{equation}
These functions must vanish for $E$ being an 
eigenvalue of the Hamiltonian (\ref{eq:RSH2}),
i.e.\ their zeros at, e.g.\ $z=0$, $G_\pm(E;0)=0$,
determine the energy eigenvalues $E$
of the {\it regular} spectrum.

\subsection{Exceptional spectrum}
\label{section:Espectrum} 
We have seen in the previous section that the zeros of the functions $G_{\pm}$
determine the energy eigenvalues 
of the {\it regular spectrum} 
of the quantum Rabi--Stark model.  

However, the functions $G_{\pm}$ have poles at certain discrete values of the energy
$E$.
Thus, while almost all eigenvalues belong to the regular spectrum, in order to determine the complete spectrum, 
one has to investigate also the values of $E$ where 
at least one of the 
$G_{\pm}$ functions diverges. 
These values of $E$ cannot belong to 
the regular spectrum, as this is determined by the set of zeros of the
$G_{\pm}$ functions. 

Instead, these values appear as candidates for the {\it exceptional eigenvalues}, 
which, together with 
particular combinations of the model parameters $\{\Delta, \gamma, g \}$, 
turn the indicial exponent $r_1 = (E+g^2 + \gamma \Delta)/(1-\gamma^2)=x_\gamma$ 
into a non--negative integer. 
Thus, in addition to the Frobenius solutions (\ref{eq:SE1}) and (\ref{eq:SE2})
corresponding to the indicial exponent $r_2=0$ which always belongs to
the (physical) Bargmann space, now also
the Frobenius solutions corresponding to an indicial exponent
$r_1=x_\gamma\in\mathbb{N}_0$
in (\ref{eq:SE1}) and (\ref{eq:SE2}) 
become members of the 
Bargmann space $\mathscr{B}$.

Similar to the case of the original quantum Rabi model
\cite{BraakReview}, we expect
two possibilities for the exceptional spectrum. 
This expectation is borne out by our numerical exploration of our exact solution of the Rabi--Stark model
which we report on in the next section \ref{section:spectrum}.

\section{Spectral structure}
\label{section:spectrum}
In this section, we report on our numerical procedure to extract the 
spectrum of the Rabi--Stark model and present
our numerical findings.
\subsection{Numerical procedure for the regular spectrum}
\label{section:RegularSpectrum1}
Given the formal solution of the quantum Rabi--Stark model, as derived in
section \ref{section:RegularSpectrum}, the recipe to numerically extract 
the \emph{regular part} of the energy spectrum can be summarized as follows: 
\begin{enumerate}
\item
In order to access the regular part of the spectrum in the {\it positive parity} 
sector for generic values of the model parameters $\{\Delta, \gamma, g \}$ 
(as before, always having set $\omega=1$), determine the expansion coefficients 
$\alpha_n$ and $\bar{\alpha}_n$ for $n=1,2,\ldots,N,$ from the 
recursion relations (\ref{eq:rra}) and 
(\ref{eq:rrb}) with initial conditions as given in (\ref{eq:Ini1}) and (\ref{eq:Ini2}), 
supplemented by the definitions 
$\alpha_{-2} = \bar{\alpha}_{-2} = \alpha_{-1} = \bar{\alpha}_{-1} =0$; 
\item
Insert the expressions for $\alpha_n$ and $\bar{\alpha}_n$ from (i) into (\ref{eq:SE1}) 
and (\ref{eq:SE2}) (with $r=r_2=0$)
via (\ref{eq:L}) and (\ref{eq:Lbar}) as well as (\ref{eq:T}) and 
(\ref{eq:Tbar}) and sum the first $N+1$ terms to obtain truncated 
series representations of 
$\phi^{(+)}(z)$ 
and 
$\bar{\phi}^{(+)}(z)$ 
(for book keeping purposes, now labeled as belonging to the positive parity sector);  
\item 
Refer to (\ref{eq:G}) to construct 
the corresponding $G_+$ function; 
\item
Locate the zeros (a.k.a. energy eigenvalues) 
$E_1,E_2,\ldots$ of $G_+(\Delta, \gamma, g|E;0)$. 
\end{enumerate}
The regular spectrum of the {\it negative parity} sector is obtained by 
repeating the steps (i)-(iv) above, but with the 
replacements $\Delta \rightarrow - \Delta$ and $\gamma \rightarrow -\gamma$ (and with 
$\phi(z)$ 
and 
$\bar{\phi}(z)$ 
in (ii) now labeled as
$\phi^{(-)}(z)$ 
and 
$\bar{\phi}^{(-)}(z)$ 
respectively, and with the energy eigenvalues obtained as 
the zeros of the corresponding function $G_-(-\Delta, -\gamma, g|E;0)$ in (\ref{eq:G})).
 
As long as the $G_{\pm}$ functions are reasonably well-behaved (as they 
are, if one does not venture too far into the deep strong coupling
regime $g>1$), the numerical 
root--finding can be carried out expeditiously, with stable results already for a truncation of 
the series in (\ref{eq:SE1}) and (\ref{eq:SE2}) to $N = 12$ terms. 

It is worth 
pointing out that the essential difference from the analogous protocol for obtaining the 
regular spectrum of the original quantum Rabi model \cite{Braak2011} is that the 
expansion coefficients $\alpha_n$ and $\bar{\alpha}_n$ now have to be derived from two 
{\it coupled} recursion relations, (\ref{eq:rra}) and (\ref{eq:rrb}). 
As discussed in sections \ref{section:Bargmann} and \ref{section:Frobenius}, 
this reflects the fact that the differential equations
(\ref{eq:ODEa}) and (\ref{eq:ODEb}) which determine the eigenfunctions 
$\phi^{(+)}(z)$ 
and 
$\bar{\phi}^{(+)}(z)$ (and $\phi^{(-)}(z)$ and $\bar{\phi}^{(-)}(z)$, respectively)
of the quantum Rabi--Stark model
have a more complex structure as compared
to the case of the original quantum Rabi model.

\subsection{Numerical procedure for the exceptional spectrum}
\label{section:ExceptionalSpectrum}
Let us now turn to the {\it exceptional part} of the spectrum which 
can be obtained by the following route:  
\begin{enumerate}
\item
For fixed model parameters $\{\Delta, \gamma, g\}$, 
rewrite the recursion relations 
(\ref{eq:rra}) and (\ref{eq:rrb}) in matrix form, i.e.\
\begin{equation}
\label{eq:coeffM}
\left( \begin{array}{c}
\alpha_n \\
\bar{\alpha}_n \\
\end{array} \right)\!
= D_n(E)^{-1} 
\mathbf{V}_n(E), \ \ \ n=2,3,...,
\end{equation}
where the vector $\mathbf{V}_n(E)$ is defined as
\begin{equation}
\label{eq:Vector}
\!\!\!\!\!\!\!\!\!\!\!\!\!\!\!\!\!\!\!\!\!\!\!\!\!
\mathbf{V}_n(E)\equiv
\left( \begin{array}{cc}
\bar{C}_{0n} & -\bar{K}_0 \\
-C_0 & K_{0n} \\
\end{array} \right)\!
\! \!\left(\! \begin{array}{c}
K_{1 n-1}\alpha_{n-1} \!-\! \bar{K}_1\bar{\alpha}_{n-1} \!-\!K_2\alpha_{n-2}\!-\!\bar{K}_2\bar{\alpha}_{n-2} \\
-C_1 \alpha_{n-1} \!+\! \bar{C}_{1 n-1} \bar{\alpha}_{n-1} \!-\! C_2 \alpha_{n-2} \!-\!\bar{C}_2 \bar{\alpha}_{n-2} 
\end{array}\!\right)\!\!,
\end{equation}
with
\begin{eqnarray}
& K_{1 n} \equiv (1-\gamma^2)n - K_1, & K_{0n} \equiv  
2w (1-\gamma^2)n +K_0,  \\
& \bar{C}_{1n} \equiv  (1-\gamma^2)n - \bar{C}_1,  & \bar{C}_{0n} \equiv  
2w(1-\gamma^2)n + \bar{C}_0,
\end{eqnarray}
and where we have defined the determinant 
and then used (\ref{eq:xgamma}),
\begin{eqnarray}
\label{eq:det}
D_n(E) \equiv K_{0n}\bar{C}_{0n}-\bar{K}_0C_0
=4w^2n(1-\gamma^2)^2\!\left[n-\frac{E+g^2+\gamma\Delta}{1-\gamma^2}\right]\!\!.
\end{eqnarray}
\item
Find the zeros $E_1, E_2,...$ of the determinant $D_n(E)$. 
These zeros locate the common singularities of the functions $G_{+}$ and $G_{-}$ 
since they cause a divergence of the corresponding $\alpha_n$ and 
$\bar{\alpha}_n$ coefficients in (\ref{eq:coeffM}).

\item
For each $E_j$ thus identified, determine whether it is also a zero of the vector
$\mathbf{V}_n(E)$ defined in (\ref{eq:Vector}).
If this is the case, the singularity is lifted in {\it both} parity sectors 
(since the zeros of the vector $\mathbf{V}_n(E)$ are invariant under 
$\Delta \rightarrow - \Delta$ and $\gamma \rightarrow - \gamma$), 
and $E_j$ becomes a 
{\it two--fold degenerate exceptional energy eigenvalue},
determining a crossing between a positive and a negative parity energy level. 

\item
If $E_j$ is {\it not} a zero of 
$\mathbf{V}_n(E)$,
the vanishing of $D_n(E_j)$ still makes room for $E_j$ to become an 
exceptional solution. 
This is because the vanishing of $D_n(E_j)$ corresponds to
the indicial 
exponent $r_1=(E_j+g^2 + \gamma \Delta)/(1-\gamma^2)$ becoming a positive 
integer, i.e.\ $r_1=n\in\mathbb{N}$. 
As a consequence, and as explained at the end of 
section \ref{section:Espectrum},
$E_j$ becomes a
{\it nondegenerate exceptional energy eigenvalue} in one of the parity sectors,
corresponding to the Frobenius solution now turned into a new 
physical Bargmann wave function at this particular juncture of parameters 
which turns $r_1$ into a positive integer. 
\end{enumerate}
As we have seen from the discussion
in this section, the eigenvalue spectrum consists of a continuous
part, the regular spectrum, which is interrupted or punctured by isolated points of the
exceptional spectrum.
These latter punctures are characterized by zeros of the determinant $D_n(E)$
which cause divergences of the $G_+$ or $G_-$ function.
The \emph{degenerate} exceptional points occur simultaneously in both
parity sectors and, thus,
determine the level crossing points (as will
be studied in an example in section \ref{section:LevelCrossings}).

As for the nondegenerate exceptional points,
the continuity of the energy levels as functions of any of the model parameters 
$\Delta, \gamma$ or $g$ implies that also these points
can only 
``fill out"
some
isolated punctures in the energy levels 
of either one or the other parity sector.
Their locations are, thus, not immediately 
visible in a numerical plot of the spectrum, 
but must be calculated analytically.
Since the nondegenerate exceptional points carry no particular significance for the interpretation of the spectrum, and also, since their detailed analytical determination
is quite involved, we shall henceforth not elaborate upon these solutions.

For a discussion of the exceptional 
spectrum in the case of the original quantum
Rabi model, see \cite{MPSFull,BraakReview}; for
a discussion of the exceptional 
spectrum of a \emph{different} generalization
of the quantum Rabi model, obtained by adding an asymmetric term $\epsilon\sigma^x$,
see \cite{LiBat,LiBatAdd}.
A detailed mathematical symmetry analysis using Lie algebra representations of 
$\mathfrak{sl}_2(\mathbb{R})$ is given for the spectrum of the original quantum
Rabi model in \cite{WY} and of the asymmetric quantum Rabi model in \cite{Wakayama}.

\subsection{Level crossings}
\label{section:LevelCrossings}
It is instructive to witness in detail how a level crossing emerges by the lifting of 
a singularity in the $G_{\pm}$ functions. 
This is but one of the advantages of the $G$ function 
approach pioneered by Braak \cite{Braak2011}: 
It allows for a compact 
encoding of the key features of the energy spectrum. 

\begin{figure}
\begin{center}
\includegraphics[width=0.77\textwidth]{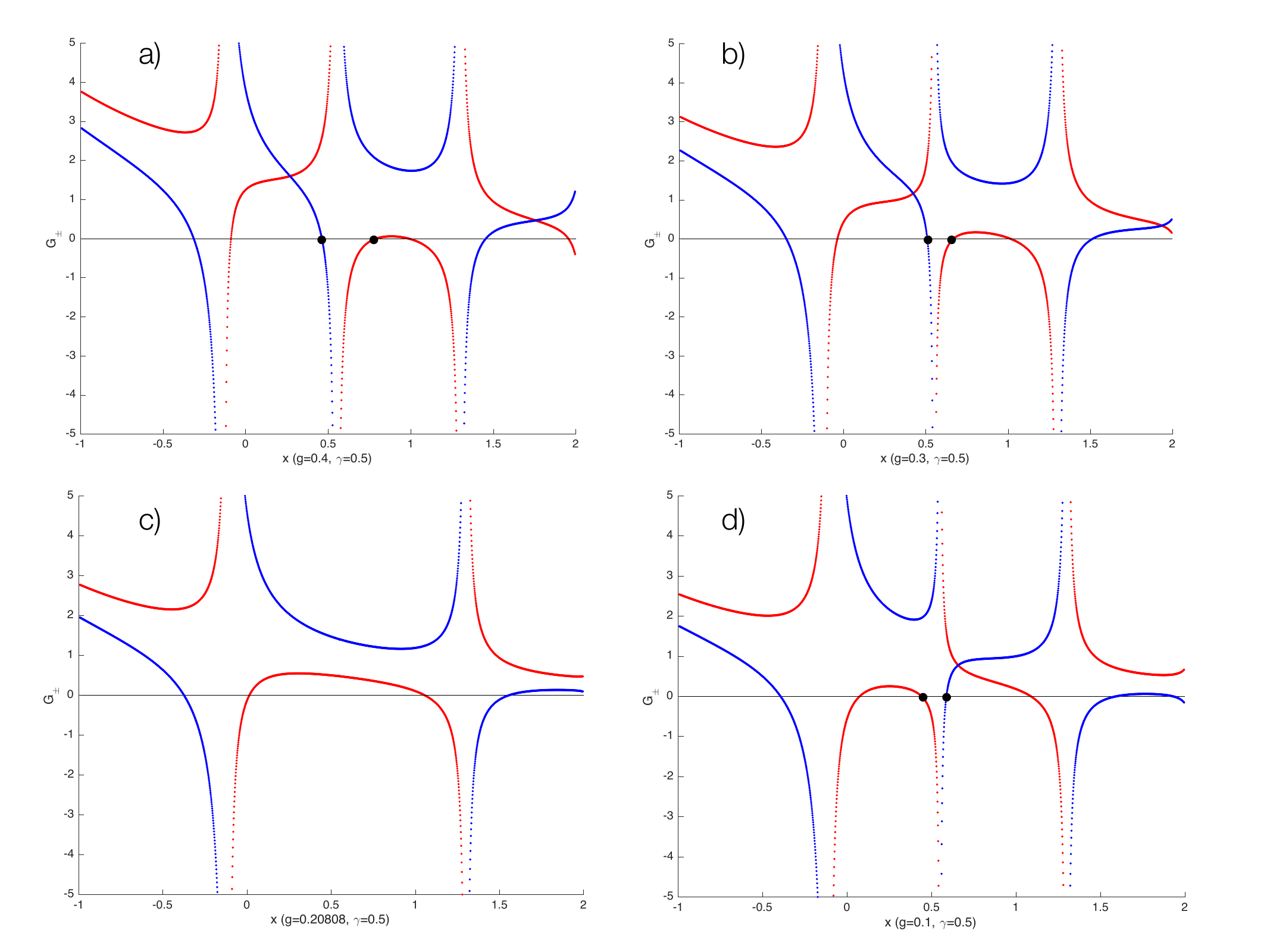}
\caption{Plots of the $G_{\pm}$ functions vs $x=E+g^2 \in [-1,2]$ for $\gamma=0.5$, 
$\Delta=0.4$ and a) g=0.4, b) g=0.3,
c) g=0.20808, and d) g=0.1. The black dots indicate the zeros of the corresponding $G_{\pm}$ functions 
closest to the singularity at $x=x_s=0.55$. In panel c) this singularity is lifted.}
\label{fig:FIG2}
\end{center}
\end{figure}

Figure \ref{fig:FIG2} exhibits a case study, where
$G_+$ ($G_-$) is shown in red (blue) versus $x=E+g^2$ in the interval $[-1,2]$
(with $E$ a running parameter which takes energy eigenvalues when $x$ becomes 
a zero of the corresponding $G_{\pm}$ function). 
The different panels correspond to different values of $g$, all with 
$\Delta= 0.4$ and $\gamma= 0.5$. 
In all panels,
the two zeros closest to the singularity at $x=x_s\approx0.55$ are marked 
with black circles. 
In the upper left panel a), the red (blue) zero is seen to be to the right (left) of  $x_s$. 
As $g$ decreases, the two zeros creep closer to $x_s$, panel b), to 
eventually coalesce 
and annihilate at $x_s$ for a value of $g=g_s$
at which the singularity 
gets lifted, panel c). 
By further decreasing $g$, the zeros move away from $x_s$, 
which has now regained its role as a locus of a singularity in $G_{\pm}$.
As seen in panel d), the zeros have traded their relative positions.

To sum up, the zeros of the $G_+$ and $G_-$ functions trade places as
$g$ is varied
across a common singularity of the two functions by lifting the singularity. 
As a consequence, a crossing between the positive and negative parity energy levels
develops at $E_{\mathrm{cross}}=x_s-g_s^2$. 
We should add that while the loci of the $G_{\pm}$ singularities in the original quantum
Rabi model appear at integer values of $x$, the loci 
for the quantum Rabi--Stark model now depend on the Stark coupling $\gamma$,
with their presence being conditioned by the vanishing of the determinant (\ref{eq:det}).

\subsection{Spectral structure of the quantum Rabi model}
\label{section:QRS}
\begin{figure}
\begin{center}
\includegraphics[width=0.85\textwidth]{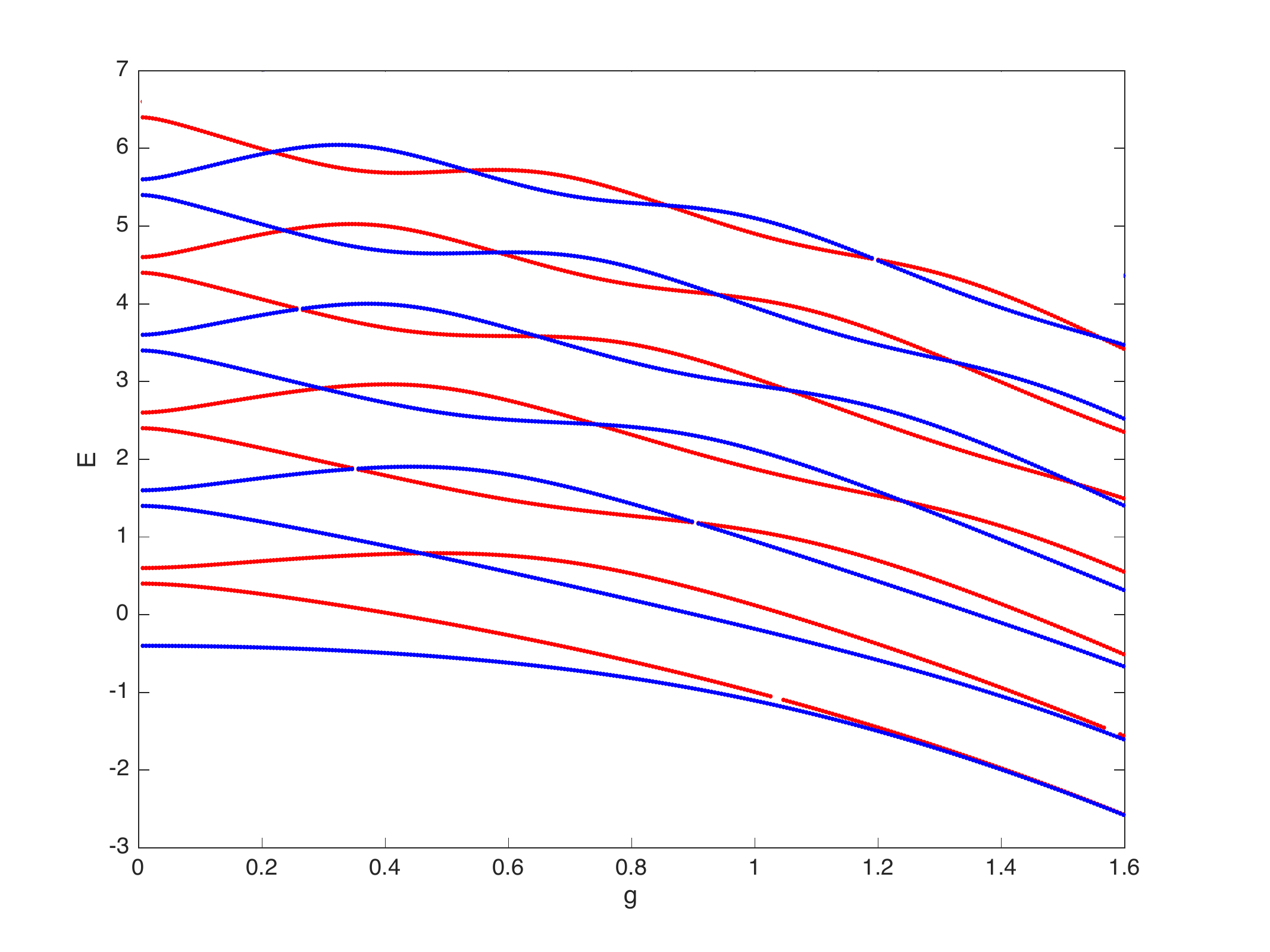}
\caption{The fourteen lowest levels in the spectrum of the quantum Rabi model $(\gamma = 0)$ for $g \in [0, 1.6]$ ($\Delta=0.4$). Red (blue) levels
correspond to the positive (negative) parity sector. The plot is composed by a dense set of points $E = x_0-g^2$ extracted
from the zeros $\{x_0\}$ of the $G_{\pm}$ functions. The glitches in some of the levels reflect that some of the zeros are
hard to resolve numerically at the level of precision used: Some zeros come extremely close to a singularity, or to a local 
extremum of a $G_{+}$  or $G_{-}$ graph which grazes the $x$-axis 
(cf. figure \ref{fig:FIG2}).}
\label{fig:Rabi}
\end{center}
\end{figure}
Before we present our numerical results for the spectrum
of the quantum Rabi--Stark model, let us set the stage 
by recalling the main characteristics of the original quantum Rabi spectrum 
\cite{Braak2011,BraakReview}. 
A low--lying part of the spectrum with the levels as function of the Rabi
coupling $g$ is depicted in figure \ref{fig:Rabi}, 
here with $g$ ranging continuously from the Jaynes-Cumming limit, 
$0<g \ll \Delta < 1$, into the opening deep strong coupling regime, $1<g <1.6$, with 
the splitting of the two--level system $\Delta =0.4$. 

The most notable feature in figure \ref{fig:Rabi}
is the absence of crossings between energy 
levels of the same parity. 
This allows for a unique labeling of the corresponding eigenstates, using the pair of 
quantum numbers $p$ and $n$,
with $p =\pm1$ denoting 
the eigenvalues of the parity
operator ${\mathcal P}$, (\ref{eq:ParityOp}), and with 
$n=0,1,2,...$ indexing the progression of 
levels of increasing energy, identified as the 
zeros of  $G_{\pm}$. 
According to the criterion proposed by Braak \cite{Braak2011}, 
the quantum Rabi model is quantum integrable because 
the eigenstates can 
be 
uniquely
identified by using {\it two} quantum numbers ($p$ and $n$),
equal to the number of degrees of freedom 
of the system 
(one two--dimensional degree of freedom characterizing the states of the 
two--level system, one 
infinite-dimensional degree of freedom for the quantum oscillator). 

Since crossings,
corresponding to the two--fold degenerate exceptional solutions 
(cf.\ figure \ref{fig:Rabi}),
appear only between levels of different parity,
one may find the resulting non--violation of the Wigner-von Neumann 
non-crossing rule \cite{LandauLifschitz} surprising: 
Quantum integrable systems are believed to violate the non-crossing rule \cite{Lieb,Jain}. 
However, as expounded in \cite{Yuzbashyan}, crossings between levels 
belonging to the same invariant subspace of a symmetry group 
(here: $\mathbb{Z}_2$ with positive and negative parity subspaces) 
are inevitable only for quantum integrable 
Hamiltonians where the number of local conserved 
quantities which depend linearly on the control 
parameter (here: the Rabi coupling
$g$) 
is maximal, i.e.\
equal to the total number of constants of motion. 
Given that the quantum Rabi model does not belong to this class, there is no 
contradiction with the criterion suggested by Braak \cite{Braak2011}. 

As seen in figure \ref{fig:Rabi}, 
levels of different parity with the same $n$ cross $n$ times before coalescing 
into near-degenerate levels for large Rabi--coupling $g$. 
This feature, present when $0<\Delta<1$, is also known from an analysis of the 
two-fold degenerate exceptional solutions, being of ``Juddian" type \cite{Judd} and 
accessible analytically \cite{KusLewenstein}. In contrast, when $\Delta >1$, 
levels of opposite parities disentangle for small and intermediary values of $g$, with at most
avoided level crossings remaining \cite{EJ}.

Given the analytical solution for the two-fold degenerate exceptional levels when $0<\Delta<1$
\cite{KusLewenstein}, one may further infer how two neighboring 
levels of different parity coalesce into a near-degenerate band \cite{Kus}: 
For a given $n$ and for large $g \gg \Delta$, the levels will tend to the 
curve $E=n-g^2$, corresponding to one of the two--fold degenerate levels 
of the quantum Rabi model with $\Delta = 0$. 
This behaviour is also easily read off from 
figure \ref{fig:Rabi}. 
It has a simple explanation: 
The two--fold degeneracy at $\Delta=0$ reflects the presence of a parity--flip symmetry:
When $\Delta=0$, the quantum Rabi Hamiltonian (\ref{eq:RH2}) 
commutes with the parity-flip operator $\sigma_x$. 
This symmetry is destroyed when turning on $\Delta$, and thus, 
the two-fold degenerate levels get split. 
However, as the 
Rabi term $\sim g$ starts to dominate the level splitting $\Delta$ of the two--level
system,
there is a smooth crossover to the two--fold degenerate level with an emergent ``approximate" parity-flip symmetry for very large $g$ 
(``approximate" in the sense that the residual terms which remain after commuting the 
Rabi Hamiltonian with the parity--flip operator $\sigma_x$ are small).
\begin{figure}
\begin{center}
\includegraphics[width=0.90\textwidth]{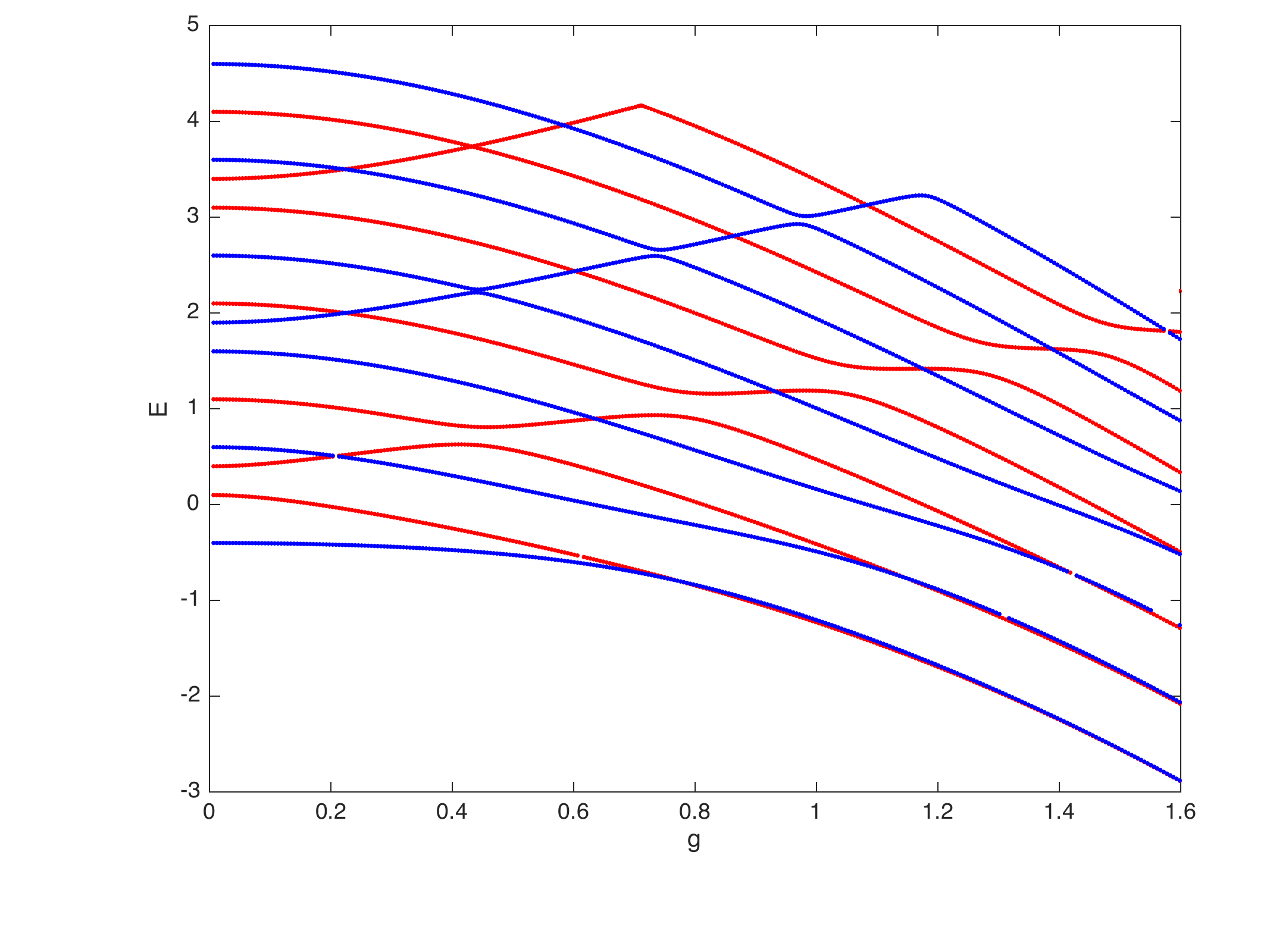}
\caption{The fourteen lowest levels in the spectrum of the quantum Rabi-Stark model with $\gamma = 0.5$ for $g \in [0, 1.6]$ ($\Delta=0.4$). Red (blue) levels
correspond to the positive (negative) parity sector. Similar to the quantum Rabi spectrum in 
figure \ref{fig:Rabi}, the glitches in some of the levels reflect 
that some of the zeros are difficult to resolve numerically at the level of precision used (cf. caption to figure \ref{fig:Rabi}) \cite{CorrigendumFig4}.} 
\label{fig:RabiStark}
\end{center}
\end{figure}
\subsection{Spectral structure of the quantum Rabi--Stark model}
\label{section:QRSS}
With the description of the original quantum Rabi spectrum as a backdrop,
we now turn to the quantum Rabi--Stark model, defined by the
Hamiltonian (\ref{eq:RSH2}).
Its fourteen lowest energy levels for $\Delta=0.4$ 
are shown in figure \ref{fig:RabiStark} \cite{CorrigendumFig4} as functions 
of $g$ in the interval  $0 \le g \le 1.6$ for $\gamma = 0.5$ and $\Delta=0.4$. 

Higher energy levels as well as spectra for larger values of 
$\Delta, \gamma$ or $g$ can also be extracted from the series 
representations of the $G_{\pm}$ functions in (\ref{eq:G}). 
However, the proliferation of singularities in the $G_{\pm}$ functions
and the slowdown of the 
convergence of the series in (\ref{eq:SE1}) and (\ref{eq:SE2}) in 
these cases make the numerics more costly. We here confine our 
attention to the chosen parameter and energy regime in figure \ref{fig:RabiStark} \cite{CorrigendumFig4}.

\begin{figure}
\begin{center}
\includegraphics[width=0.80\textwidth]{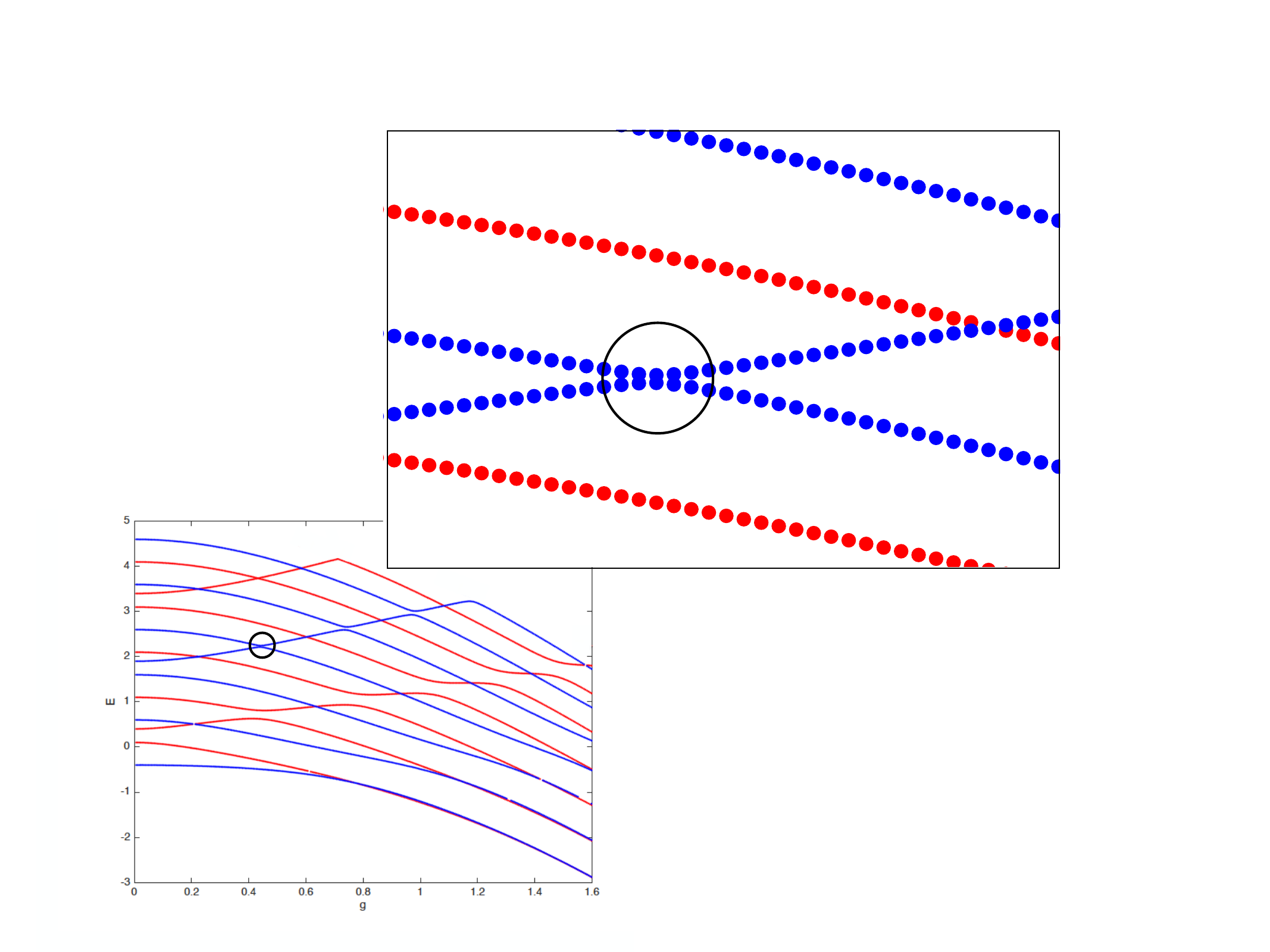}
\caption{Zoom in of the spectrum of the quantum Rabi-Stark model in figure 
\ref{fig:RabiStark}, showing an avoided level crossing 
at $g\approx 0.5$.}
\label{fig:AvoidedCrossing}
\end{center}
\end{figure}

Inspection of the spectrum in figure \ref{fig:RabiStark} \cite{CorrigendumFig4} shows that crossings 
of energy levels of the same parity remain absent in the 
presence of the added nonlinear Stark coupling term. 
What may first appear as equal--parity level crossings 
(e.g.\ between the fifth and sixth blue curves close to $g=0.5$ 
in figure \ref{fig:RabiStark} \cite{CorrigendumFig4}), at close scrutiny are revealed 
to be {\it avoided level crossings}, cf. figure \ref{fig:AvoidedCrossing}.
Intriguingly, by tuning the splitting $\Delta$ of the two-level system, the avoided 
level crossings can be made progressively sharper, suggesting
the possibility of a nonanalyticity for a critical value of $\Delta$, 
cf. figure \ref{fig:RabiStark} \cite{CorrigendumFig4}. 

While it is tempting to
speculate that this incipient nonanalyticity may be a precursor of an 
``excited state quantum phase transition" \cite{Heiss,Leyvraz,Cejnar},
this would be premature.
In order to present support for such a transition, one must first and foremost 
establish a critical energy below which there is a symmetry breaking, 
with the critical energy accompanied by a singularity in the density of states. 
Let us note in passing that Puebla {\it et al.}, using an effective Hamiltonian,
have recently conjectured 
that such a transition 
may actually be present
in the original quantum Rabi model \cite{Puebla} (see also \cite{LarsonIrish}).
Their approach was very recently generalized \cite{Shen} for an anisotropic
quantum Rabi model where the rotating and counterrotating parts of the
Rabi coupling term acquire different coupling strengths:
$g\left(\sigma^++\sigma^-\right)\left(a+a^\dagger\right)
\rightarrow g_r\left(\sigma^+a+\sigma^-a^\dagger\right)
+g_{cr}\left(\sigma^+a^\dagger+\sigma^-a\right)$. 
We further note, again in passing, that this anisotropic model is also within the reach
of the experimental proposal of Grimsmo and Parkins \cite{GP2013},
that it admits an exact solution
and that it can be used in a variety of physical situations \cite{Xie2,Tomka,Zhang},
including a proposed realization of supersymmetry \cite{Tomka2}.
In view of these developments, it will be very interesting to examine the 
anisotropic generalization of the Rabi--Stark model.

As for the avoided level crossings in the quantum Rabi--Stark model, 
we expect that they rather reflect the model's integrability 
(in the sense of Braak \cite{Braak2011}): 
In order to uphold integrability as the original quantum Rabi levels 
get reshuffled
by the nonlinear Stark term with strength $\gamma$,
same--parity avoided level crossings appear in various parts of the spectrum. 
If same-parity level crossings had developed,
this would have required the model to be "superintegrable" 
\cite{CauxMossel} for 
nonzero values of $\gamma$, supporting an
additional "good quantum number" by which the energy levels could be uniquely labeled.
This $-$ by itself quite unlikely $-$ scenario is made the more improbable by 
the presence of 
the avoided level crossings in figure \ref{fig:RabiStark} \cite{CorrigendumFig4}. 

As is evident from figure \ref{fig:RabiStark} \cite{CorrigendumFig4}, 
the reshuffling of levels as $\gamma$ increases also leads to more densely spaced levels.
This latter ``compression" effect is anticipated from the trivial solution
of the Rabi-Stark model with Rabi coupling $g=0$ 
which is solvable by elementary means since
for this case the Hamiltonian is diagonal.
Explicitly, the eigenvalues for $g=0$ are
\begin{equation}
\label{eq:G0Levels}
E^{\pm}_n(g=0,\gamma)=(1\pm\gamma)n\pm\Delta\quad n=0,1,2,\ldots
\end{equation}
from which it can be seen that more and more levels accumulate at 
$E^-_n(g=0,\gamma\rightarrow1)\rightarrow-\Delta$
as $\gamma\rightarrow1$.
In the same limit $\gamma\rightarrow1$, 
the other levels
become equally spaced, 
$E^+_n(g=0,\gamma\rightarrow1)\rightarrow\Delta+2n$, starting from $+\Delta$.
Please note that the label $\pm$ in (\ref{eq:G0Levels}) does not refer to parity.
The parity of these levels is determined by the parity operator (\ref{eq:ParityOp}).

In this context: We have already dispelled a possible concern 
about nonviolations of the Wigner--von Neumann no--crossing rule. 
But what about the Berry--Tabor criterion \cite{BerryTabor} that an 
integrable model exhibits a Poissonian distribution of energy levels? 
Similar to the original quantum Rabi model, the levels for the quantum 
Rabi--Stark model when
$\gamma \neq 0$ appear to be distributed fairly regularly 
(cf. figure \ref{fig:RabiStark} \cite{CorrigendumFig4}) and not Poissonian.
Thus both, the quantum Rabi and Rabi--Stark spectral distributions
fail this test of integrability. 
However, it is important to be precise about the range of 
applicability of the Berry--Tabor criterion:  
It has been proved only in the 
semiclassical limit, and moreover assumes that the theory 
supports only continuous degrees of freedom \cite{BerryTabor}. 
None of this applies to the quantum Rabi or Rabi--Stark model.

Figure \ref{fig:RabiStark} \cite{CorrigendumFig4} reveals that the level crossings of opposite--parity
levels for the Rabi--Stark model
($\gamma \neq 0$) no longer follow the simple 
``braiding rule" of the Rabi model
where two neighbouring levels with quantum number 
$n$ cross $n$ times. 
As a case in point, when $\gamma=0.5$ (figure \ref{fig:RabiStark} \cite{CorrigendumFig4}), the first  
four pairs 
of opposite-parity levels cross at most two times before coalescing 
into a near-degenerate level.
The implied reduction of two--fold degenerate exceptional solutions 
of the differential equations (\ref{eq:ODEa}) and (\ref{eq:ODEb}) when $\gamma \neq 0$, 
$-$ underlying the reduction of opposite--parity level crossings $-$
should have an explanation in terms of the $\gamma$-dependent loci of the singularities
in the $G_\pm$ functions (cf. the discussion in section \ref{section:ExceptionalSpectrum}). 
However, to pinpoint the resulting structure of level crossings in the spectrum goes 
beyond the aim of this work. Indeed, a closer analytic examination of the $G_\pm$ functions 
remains a challenge for the future.

\begin{figure}
\begin{center}
\includegraphics[width=0.8\textwidth]{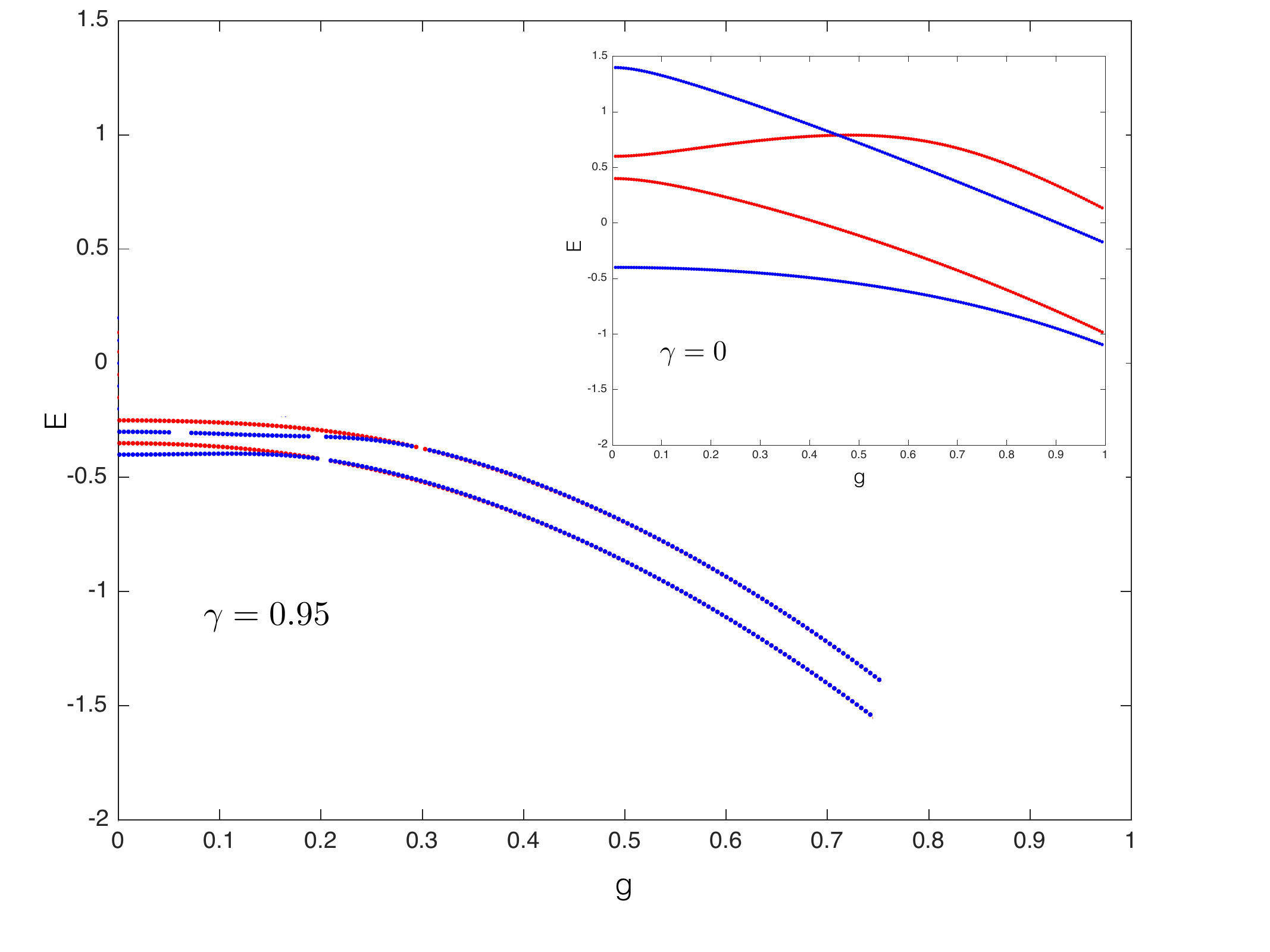}
\caption{The four lowest energy levels of the quantum Rabi-Stark model with $\gamma=0.95$ in the interval
$g \in [0,0.75]$. For comparison, the four lowest levels of the quantum Rabi model ($\gamma=0$) are shown
in the inset for $g \in [0,1]$.
In both cases $\Delta=0.4$ \cite{CorrigendumFig6}.}
\label{fig:Degeneracy}
\end{center}
\end{figure}

As already mentioned, there occurs a compression of the spacings of the
energy levels as the coupling $\gamma$ of the nonlinear Stark term increases.
This effect may facilitate $-$ but does not explain $-$ that for large $\gamma$, 
neighbouring levels with opposite parity coalesce 
into near--degenerate levels 
already for quite small values of $g$. 
For an example, see figure \ref{fig:Degeneracy} \cite{CorrigendumFig6}, where the two lowest pairs of 
opposite--parity levels are shown as functions of $g$ when $\gamma = 0.95$ 
(with the two lowest pairs of levels for $\gamma=0$
shown for comparison in the inset).  Already for $g = 0.6$ in the figure, the difference 
between the two lowest levels is $< 0.00004$ and then rapidly decreases as $g$ increases further.
The rapid approach in figure \ref{fig:Degeneracy} \cite{CorrigendumFig6}
to near--degeneracy of the energy levels for $\gamma=0.95$, 
as compared to the case with $\gamma=0$, in fact is
surprising in view of the Hamiltonian ${\cal H}$
in (\ref{eq:RSH2}). 
In order to make ${\cal H}$
approximately invariant under 
a parity flip with parity--flip operator $\sigma_x$
$-$ which 
guarantees near--degeneracy of neighbouring levels of opposite parity $-$ 
the Rabi term must now dominate both the 
term of the two--level system (with splitting $\Delta$) {\it and} the nonlinear 
Stark term (with coupling parameter $\gamma$).
When $\gamma > \Delta$, as in figure \ref{fig:Degeneracy} \cite{CorrigendumFig6}, one would 
expect that the approximate parity--flip symmetry, and thus, 
the concurrent near--degeneracy, would set in for values of 
$g$ larger than what is required when $\gamma=0$. 
However, as revealed by the same figure (with its inset), the opposite is the case! 

These numerical observations suggest 
that the spectral compression dramatically enhances the effect of a reduced parity-flip 
symmetry breaking (so as to boost the early onset of near-degenerate energy levels), or else, that some hidden symmetry 
is at play for values of $g$ for which the parity--flip symmetry is still manifestly broken. 
While we find this latter alternative to be rather unlikely, 
we should alert the reader that there are indeed claims 
that already the original quantum Rabi model has a hidden 
symmetry, with implications for its dynamics \cite{Gardas}. 
In any event, a further analysis of the compression of the spectrum  
for increasing $\gamma$, with the concurrent rapid emergence of 
near--degeneracy, seems to be called for.

\section{Discussion and summary}
\label{section:DS}
The generalized quantum Rabi model described by the Hamiltonian (\ref{eq:RSH2}),
the quantum Rabi--Stark model,
is particularly interesting from the point of view that it offers a further
tunable parameter, the Stark coupling $\gamma$, which can be used to investigate
various regimes of the model which may be less accessible for the original quantum Rabi model
where the Stark coupling vanishes.
This is an especially intriguing aspect of the model since an experimental realization
has been proposed with all the energy parameters $\{\omega,\Delta,\gamma,g\}$ freely
and independently variable \cite{GP2013}.

In the investigation reported here, the quantum Rabi--Stark model
has been shown to be exactly solvable and also quantum integrable in the sense
of quantum integrability introduced by Braak in his seminal work on the original
quantum Rabi model \cite{Braak2011}. 
Furthermore, we have obtained the exact analytical solution of the 
quantum Rabi--Stark model Hamiltonian adapting the methods devised for the 
original quantum Rabi model in \cite{Braak2011}.
In particular, we highlighted the differences created by the nonlinear Stark term 
$\gamma\sigma^za^\dagger a$ in the generalized model.
One of these differences concerns the reproduction of the results for the original
quantum Rabi model from those of the generalized model.
The naive limit $\gamma\rightarrow0$ fails for the 
non--zero indicial exponents of the Frobenius
solution of the generalized model.
This observation will be crucial for a study of the exceptional points in the spectrum of the
quantum Rabi--Stark model.

From the exact solution, we constructed functions $G_\pm(E;z)$
which can be used to numerically extract the regular part of the spectrum of the model.
The exact solution 
also allows for a classification of the exceptional part of the spectrum which consists
in its turn of a degenerate and a nondegenerate part.
The spectrum and its properties, especially its dependence on the parameter
$\gamma$ of the non--linear Stark term in the Hamiltonian has been the major aim of the
investigation we have reported on in this paper. 
As detailed in the last section,
the low-lying Rabi-Stark spectrum in the experimentally most relevant ultrastrong and opening deep strong 
regimes of the Rabi coupling exhibits two striking features absent from the original quantum Rabi spectrum:
Distinctive avoided level crossings within each parity sector, and the onset of two-fold near-degenerate 
levels already in the ultrastong regime when $\gamma$ becomes sufficiently large. 
While the same-parity
avoided level crossings most likely reflect the integrabiltiy of the model $-$ as bolstered 
by the underlying $\mathbb{Z}_2$ 
parity symmetry \cite{Braak2011} $-$ the rapid onset of near-degeneracy remains 
more intriguing. 
To provide for its 
interpretation or explanation remains an open problem.

\ack 
We thank Murray Batchelor, Daniel Braak, 
Mang Feng, and Michael Tomka for
illuminating discussions, and Arne Grimsmo and Jonas Larson
for valuable correspondence.
Furthermore we thank Elinor Irish for pointing out an erroneous statement
in an early version of the manuscript.
This work was supported by
STINT (Grant No. IG2011-2028) and the Swedish Research 
Council (Grant No. 621-2014-5972).

\newpage

\section{Corrigendum (published in J. Phys. A: Math. Theor.
{\bf 56}, 345302 (2023))}

In our paper (pages 1-25; published in J. Phys. A: Math. Theor. {\bf 50} 294004 (2017)), 
we solved exactly the quantum Rabi-Stark model, a
generalization of the quantum Rabi model
(for a recent elementary discussion of the quantum Rabi model, see \cite{Book}), 
and discussed its spectral structure.
Doing this required a generalization of Braak's method \cite{Braak2011} developed
for the exact solution of the quantum Rabi model.
This method crucially uses the $\mathbb{Z}_2$  symmetry of the models.
Working in the Bargmann space representation, the ensuing coupled first-order singular
ordinary differential equations are solved by a modified power series ansatz,
the so-called Frobenius method.
The coefficients in this power series are then given, for the quantum Rabi-Stark model,
by coupled three-term recurrence relations.
Once these recurrence relations are solved numerically, the energy spectra
are obtained as the zeros of the so-called $G$ functions.

The recurrence relations, equations (47) and (48), for the power series coefficients contain twelve coefficient
functions, equations (33)-(44) in our original paper,
which depend on the model parameters (i.e. the oscillator frequency
($\omega\equiv1$), the qubit splitting
($\Delta$), the Rabi ($g$), and Stark ($\gamma$) coupling) and the energy $E$.

 \begin{figure}[b]
\begin{center}
\includegraphics[width=0.5\textwidth]{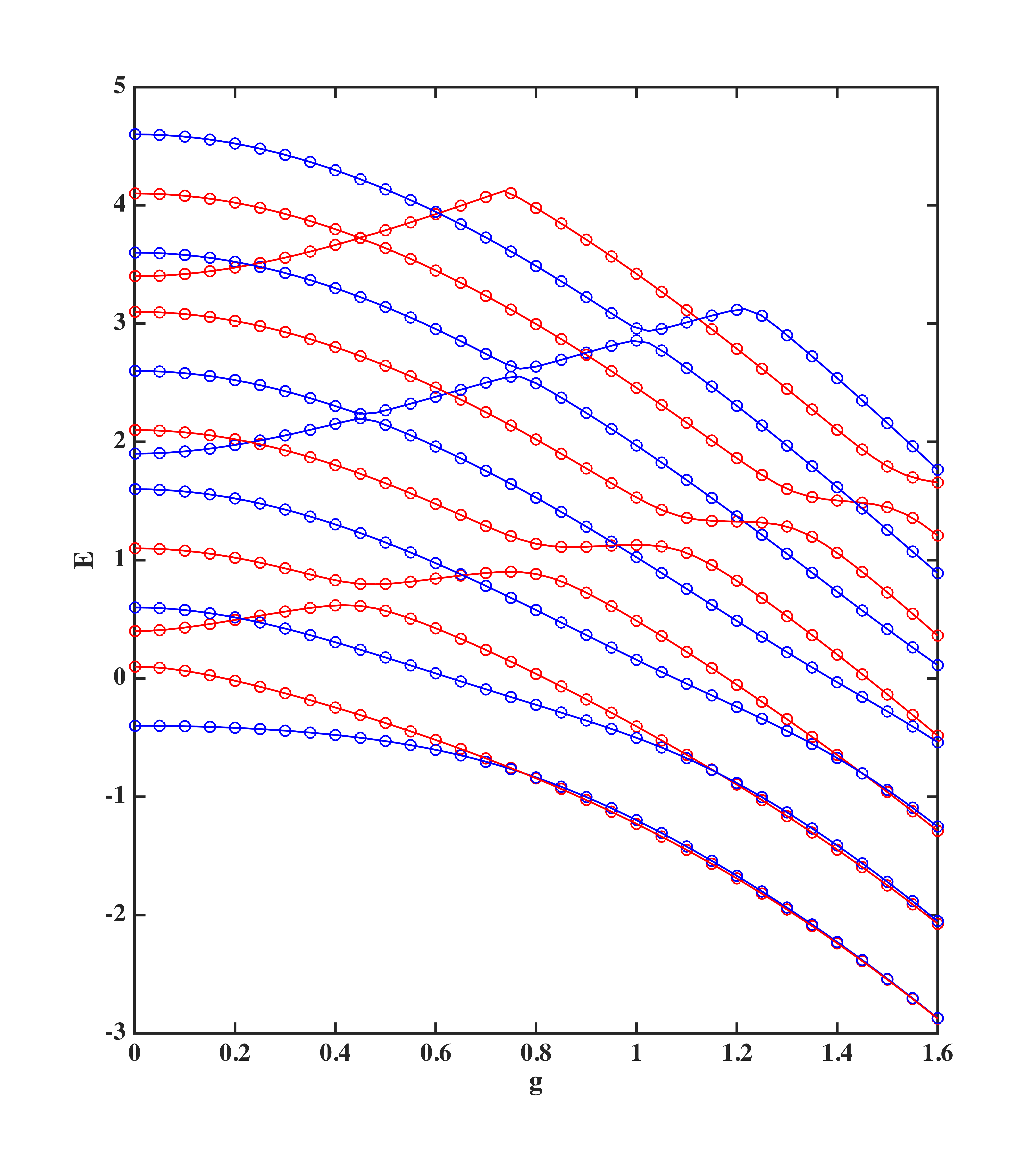}
\caption{Energy spectra of the quantum Rabi-Stark model for $\gamma=0.5$
              ($\Delta=0.4$).
              The blue (red) lines correspond to odd (even) parity levels calculated by
              exact numerical diagonalization (truncation number $N=400$).
              The blue (red) rings mark odd (even)
               parity levels obtained from
              the coupled recurrence relations (47), (48).
                 } 
\label{fig:Fig4Compare20230529}
\end{center}
\end{figure}

Two of these coefficients, $\bar{C}_2$ in equation (39) and $\bar{C}_1$ in equation (40)
were miscalculated and should be replaced by
\begin{eqnarray}
\label{eq:c1}
\bar{C}_2 &=&(1-\gamma^2)w+g=g(\sqrt{1-\gamma^2}+1),\\
\label{eq:c2}
\bar{C}_1&=&E-2gw-g^2+\gamma\Delta ,
\end{eqnarray}
 where $w=g/\sqrt{1-\gamma^2}$.

\begin{figure}
\begin{center}
\includegraphics[width=0.5\textwidth]{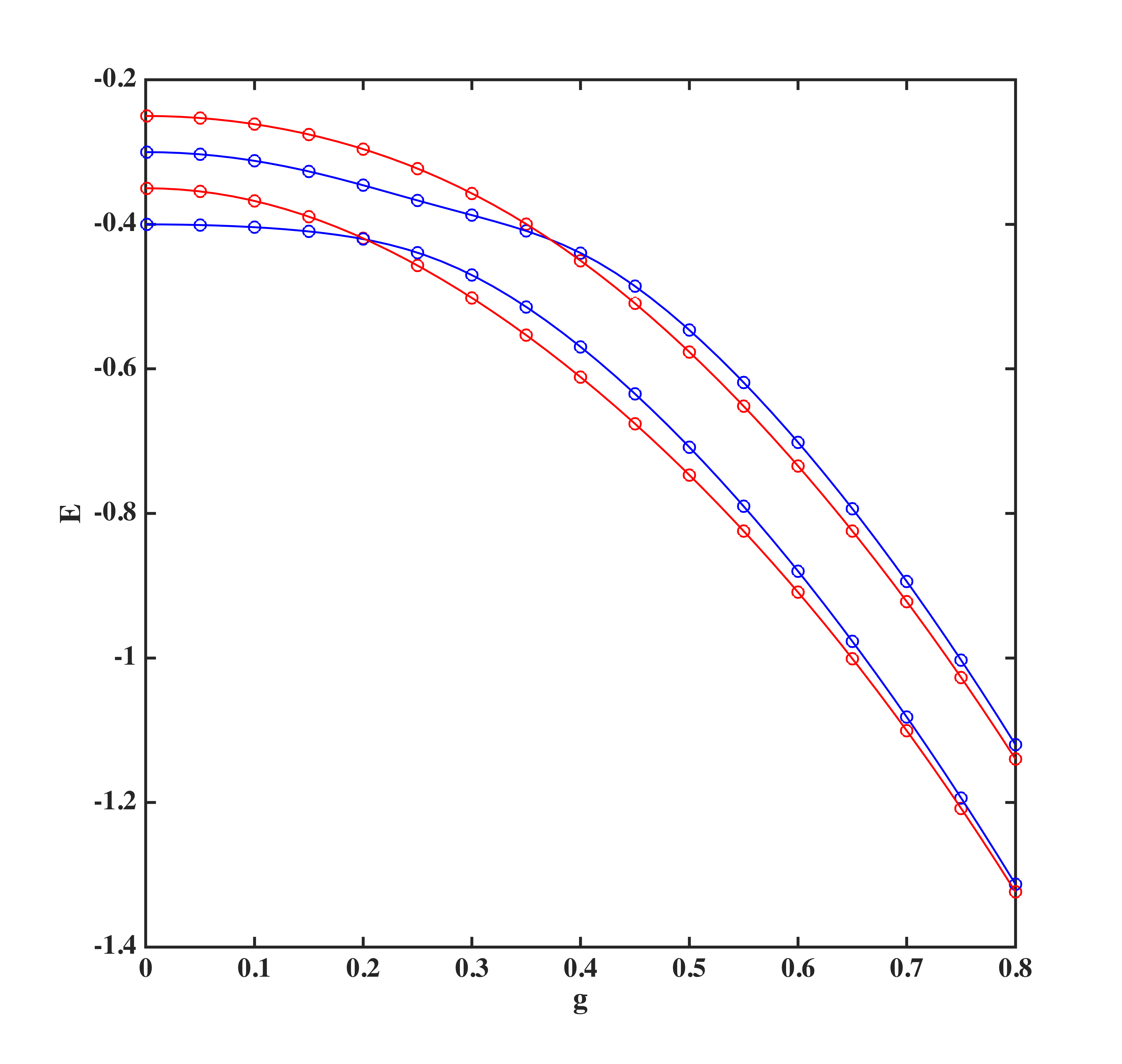}
\caption{Energy spectra of the quantum Rabi-Stark model for $\gamma=0.95$
              ($\Delta=0.4$).
              Truncation number and colour and symbol code are the same as in 
              figure \ref{fig:Fig4Compare20230529}.
                 } 
\label{fig:Fig6Compare_20230527}
\end{center}
\end{figure}
 
Using the corrected coefficients (\ref{eq:c1}) and (\ref{eq:c2}), one obtains the
energy spectra in figures \ref{fig:Fig4Compare20230529} and \ref{fig:Fig6Compare_20230527}
(replacing figures 4 and 6, respectively, in our original paper).
Also shown in figures \ref{fig:Fig4Compare20230529} and \ref{fig:Fig6Compare_20230527} are comparisons with energy spectra obtained
by exact numerical diagonalization with a truncation number of $N=400$.
The numerically exact diagonalization of the quantum Rabi-Stark model is
completely independent of the procedure carried out in our original paper.
However, the fit of the spectra obtained by these two independent methods is
excellent.

While the energy spectra in figures 4, 5 and 6 of our paper are numerically incorrect
and should not be used for extracting quantitative information, the 
general conclusions drawn in the paper remain unchanged and valid.
This includes the scenario for the lifting of degeneracies in the $G$ functions
illustrated in figure 2 of our original paper.

\ack
Our special thanks go to Lei Cong for performing the numerical calculations for this
corrigendum, in particular the exact numerical diagonalization results.
Furthermore, we thank Daniel Braak, Lei Cong, and Elinor Irish for many useful
discussions.

\end{document}